\documentclass[12pt]{article}

\usepackage{physics} 
\usepackage{siunitx} 
\usepackage{enumerate} 
\usepackage{pgfplots}
\usepackage{pgfplotstable}
\usepackage{tikz,pgfplots}
\usepackage{amsmath}  
\usepackage{wasysym} 
	
\usepackage{geometry}
\usepackage{authblk}
\usepackage{tensor}
\usepackage{braket}
\usepackage{stackengine}
\usepackage[backend=biber,style=numeric-comp,backref=true,sorting=none]{biblatex}

\renewbibmacro{in:}{}
\DeclareFieldFormat{pages}{#1}
\usepackage{hyperref}

\pgfplotsset{compat=1.14}

\begin{document}

\title{Spin density matrix for high-spin particles} 
\author[1]{Shaogang Peng}
\author[1]{Xin Chen\thanks{\href{mailto:xchen22@mail.tsinghua.edu.cn}{xchen22@mail.tsinghua.edu.cn}}}
\author[1]{Yuxuan Zhang}
\author[1]{Qi Cai}
\author[1]{Naiwen Xing}
\author[2]{Yue Xu\thanks{\href{mailto:yue.xu@cern.ch}{yue.xu@cern.ch}}}
\affil[1]{Tsinghua University}
\affil[2]{University of Washington}

\date{} 

\maketitle  

\begin{abstract}
The spin density matrix (SDM) characterizes the spin state of a particle, which provides important information for understanding its production mechanism. A general discussion of the SDM under parity conservation, and specific SDM parameterizations for spin-1 and spin-2 particles are provided, which respect all the expected constraints. The imaginary (real) part of SDM is related to the degree of odd-rank (even-rank) polarization, with specific expressions given for spin-1 and spin-2 cases. 
Examples of applications of the SDM to angular analyses of several decay processes with different spin-parity hypotheses are also provided, which could be helpful in experimental data analyses.
\end{abstract}

Angular analyses in particle collisions can measure the polarization state of the decaying resonance, determine its spin and parity quantum numbers, provide information on different partial waves, and reveal production mechanisms or new resonances in decay processes~\cite{Jacob:1959at,Chung:1971ri,Schilling:1969um,Arifi:2020ezz,BESIII:2020lkm,Ringland:1968vdb,Gottfried:1964nx,Irving:1974ak}. 
The spin density matrix (SDM) is an integral part of the angular analysis. It contains complete information about the spin state of a single particle at production~\cite{Fano:1957zz,Leader:2001nas,Boudjema:2009fz}. 
When the particle is correlated with other particles in a larger quantum system, a reduced SDM is used if the remaining degrees of freedom are not observed~\cite{Fano:1957zz,Leader:2001nas}. 
There are many studies of SDMs for spin-1~\cite{BIEDENHARN1958104,Schilling:1969um,Bourrely:1980mr,Bacchetta:2000jk,Liang:2004xn,Chen:2020pty} and spin-3/2~\cite{Song:1967vpt,Kim:1976dn,Choi:1989yf,Zhao:2022lbw,Zhang:2023wmd} particles, whereas relatively few studies have addressed spin-2 particles~\cite{Boudjema:2009fz}. 
We present a general formalism for the spin-1 and spin-2 SDM and discuss several applications in collider physics, relating the SDM elements to measurable polarization observables.

Consider a resonance $R$ produced in hadron-hadron or $e^+e^-$ collisions via the process $A+B \to R+C$, where $R$ may subsequently decay into $D_1$ and $D_2$, while $C$ denotes the remainder of the final state, which may be unobserved.
A production plane is defined by the beam particles and the outgoing $R$ and $C$. 
We parameterize the SDM in the helicity frame~\cite{Jacob:1959at,Chung:1971ri} by boosting to the rest frame of $R$, where its spin state can be described using the usual non-relativistic angular-momentum formalism. In this frame, the $z$-axis (quantization axis) coincides with the momentum of $R$ ($\boldsymbol{p}_R$) in the lab frame, and the $y$-axis  coincides with $\boldsymbol{p}_{beam}\times\boldsymbol{p}_R$, where $\boldsymbol{p}_{beam}$ is the momentum of one incident beam. The $x$-axis is naturally obtained by $\hat{y}\times\hat{z}$. 

The resonance $R$ is usually produced through strong or electromagnetic interactions, which conserve parity.
Reflection about the production plane leaves the system invariant, leading to the following relation among the SDM elements of $R$~\cite{Chung:1971ri,Schilling:1969um}:

\begin{equation}
\rho_{mm^\prime} = (-1)^{m-m^\prime} \rho_{-m,-m^\prime}
\label{eq:eq1}
\end{equation}
where $m$ and $m^\prime$ are the third components of  $R$'s spin. They are arranged in a descending order for locating the position of an element in $\rho$. Considering the constraints of hermiticity and Eq.~(\ref{eq:eq1}), the most general form of $\rho$ for a spin-2 resonance can be expressed as 

\begin{equation}
\begin{aligned}
& \rho = 
& \begin{pmatrix}
a_1 &  b_1 & b_3  & b_5  & b_6 \\
b_1^* &  a_2 & b_2 & b_4 & -b_5^* \\
b_3^* &  b_2^* & a_3 & -b_2^*  & b_3^*  \\
b_5^* &  b_4 & -b_2 & a_2 & -b_1^* \\
b_6 &  -b_5 & b_3 & -b_1 & a_1 \\
\end{pmatrix},
\end{aligned}
\label{eq:eq2}
\end{equation}
where parameters $a_{1,2,3}$ and $b_{4,6}$ are all real, and $b_{1,2,3,5}$ are all complex numbers. With the constraint $\text{Tr}(\rho)=1$ (i.e. $2a_1+2a_2+a_3=1$), there are 12 independent parameters in total. In contrast, in the ordinary case without parity conservation, the number of parameters needed for the specification of the SDM is $(2s+1)^2-1$ (i.e. 24 for $s=2$). Parity conservation significantly reduces this number. 
Finally, the SDM in Eq.~(\ref{eq:eq2}) must be positive semidefinite. Namely, its elements satisfy $|\rho_{ij}|^2\leq\rho_{ii}\rho_{jj}$ and $\rho_{ii}\geq 0$~\cite{DiSalvo:2020run}. 

The SDM can be decomposed into a set of spherical tensor operator matrices $T^L_M$, with $0\leq L\leq 2s$ and $-L\leq M\leq L$ ($L$ is called the rank of the tensor operator). Its element is

\begin{equation}
\left(T^{L}_M \right)_{mm^\prime} = \left(s m^\prime LM | sm\right),
\label{eq:eq3}
\end{equation}
whose right-hand side is just the Clebsch--Gordan coefficient. 
The corresponding expansion coefficients, referred to as multipole parameters, can be expressed as~\cite{Chung:1971ri}

\begin{equation}
t^{L*}_M = \text{Tr}\left(\rho T^{L\dagger}_M \right) = \sum_{mm^\prime} \rho_{mm^\prime} \left(s m^\prime LM | sm\right).
\label{eq:eq4}
\end{equation}
To quantify the deviation of the system from the fully mixed state described by $\rho = I/(2s+1)$, the degree of rank-$L$ polarization is defined as~\cite{Leader:2001nas}

\begin{equation}
d_L = \sqrt{\frac{2L+1}{2s} \sum_M \left| t^L_M \right|^2}.
\label{eq:eq5}
\end{equation}
For spin-2 particles, the explicit expressions for $d_L$ are

\begin{equation}
{\footnotesize
\begin{aligned}
d_1 & = \left| \sqrt{2} \text{Im}(b_1) + \sqrt{3} \text{Im}(b_2)\right|, \\
d_2 & = \sqrt{\frac{5}{14}} \sqrt{ \left(2a_1-a_2-a_3 \right)^2 + 2\left[\sqrt{6} \text{Re}(b_1) + \text{Re}(b_2) \right]^2 + \left[2\sqrt{2} \text{Re}(b_3) + \sqrt{3} b_4\right]^2 }, \\
d_3 & = \sqrt{  \left[\sqrt{3} \text{Im}(b_1) - \sqrt{2} \text{Im}(b_2) \right]^2 + 5\text{Im}(b_3)^2 + 5\text{Im}(b_5)^2 }, \\
d_4 & = \frac{1}{\sqrt{14}} \sqrt{\left(a_1-4a_2+3a_3 \right)^2 + 10\left[\text{Re}(b_1) - \sqrt{6} \text{Re}(b_2)  \right]^2 + 5  \left[ \sqrt{6} \text{Re}(b_3) - 2 b_4 \right]^2 + 70 \text{Re}(b_5)^2 +35 b_6^2 }, \\
\end{aligned}}
\label{eq:eq6}
\end{equation}
from which one immediately sees that $d_1$ and $d_3$ only depend on the imaginary part of $\rho$, while  $d_2$ and $d_4$ only depend on the real part of it. 
Notice that the degree of rank-1 polarization $d_1$ is proportional to the magnitude of the spin-polarization vector $\boldsymbol{\mathcal{P}} = \text{Tr} (\rho \boldsymbol{S}) $, where $\boldsymbol{S}$ reads

\begin{equation}
{\footnotesize
\begin{aligned}
S_x = 
\begin{pmatrix}
0 & 1 & 0 & 0 & 0 \\
1 & 0 & \frac{\sqrt{6}}{2} & 0 & 0 \\
0 & \frac{\sqrt{6}}{2} & 0 & \frac{\sqrt{6}}{2} & 0 \\
0 & 0 & \frac{\sqrt{6}}{2} & 0 & 1 \\
0 & 0 & 0 & 1 & 0 \\
\end{pmatrix},~
S_y = 
\begin{pmatrix}
0 & -i & 0 & 0 & 0 \\
i & 0 & -\frac{\sqrt{6}}{2}i & 0 & 0 \\
0 & \frac{\sqrt{6}}{2}i & 0 & -\frac{\sqrt{6}}{2}i & 0 \\
0 & 0 & \frac{\sqrt{6}}{2}i & 0 & -i \\
0 & 0 & 0 & i & 0 \\
\end{pmatrix},~
S_z = 
\begin{pmatrix}
2 & 0 & 0 & 0 & 0 \\
0 & 1 & 0 & 0 & 0 \\
0 & 0 & 0 & 0 & 0 \\
0 & 0 & 0 & -1 & 0 \\
0 & 0 & 0 & 0 & -2 \\
\end{pmatrix}.
\end{aligned}}
\label{eq:eq7}
\end{equation}
It is straightforward to obtain $\text{Tr} (\rho S_x) = \text{Tr} (\rho S_z) = 0$, and $\text{Tr} (\rho S_y)  = -2\left[ 2 \text{Im}(b_1) + \sqrt{6} \text{Im}(b_2)\right]$. 
Therefore, any nonzero polarization vector $\boldsymbol{\mathcal{P}}$ must be perpendicular to the production plane.
This is understandable because the reflection operation does not flip the spin in the $y$-direction due to its axial-vector property, but will flip the helicities of all particles in the initial and final states.

Finally, an overall degree of polarization is defined as

\begin{equation}
d = \sqrt{ \sum_{L\geq 1} d_L^2},
\label{eq:eq8}
\end{equation}
which takes into account the contributions from all $d_L$'s and lies in the range $0\leq d\leq 1$.

The most general spin-1 SDM consistent with parity conservation can be written analogously as

\begin{equation}
\begin{aligned}
& \rho = 
& \begin{pmatrix}
a_1 &  b_1 & b_2 \\
b_1^* & a_2 & -b_1^* \\
b_2 & -b_1 & a_1 \\
\end{pmatrix},
\end{aligned}
\label{eq:eq9}
\end{equation}
where $a_{1,2}$ and $b_2$ are real, and $b_1$ is complex. It is also required that $a_i\geq 0$  and $2a_1+a_2=1$. The degrees of rank-1 and rank-2 polarizations read

\begin{equation}
{\small
\begin{aligned}
d_1 & = \sqrt{6} \left|\text{Im}(b_1)\right|, \\
d_2 & = \frac{1}{2} \sqrt{ (1-3a_2)^2  + 24\text{Re}(b_1)^2 + 12 b_2^2 }. \\
\end{aligned}}
\label{eq:eq10}
\end{equation}
And the overall degree of polarization is

\begin{equation}
d = \frac{1}{2} \sqrt{ (1-3a_2)^2  + 24\text{Re}(b_1)^2 + 24\text{Im}(b_1)^2 + 12 b_2^2 }.
\label{eq:eq11}
\end{equation}
It is evident from Eq.~(\ref{eq:eq10}) that 
$d_1$ depends only on $\text{Im}(\rho)$, and a sufficient condition for $d_2>0$ is $a_2\neq 1/3$. For reference, the matrix representations of the spin-1 operators are

\begin{equation}
{\small
\begin{aligned}
S_x = 
\begin{pmatrix}
0 & \frac{1}{\sqrt{2}} & 0 \\
\frac{1}{\sqrt{2}} & 0 & \frac{1}{\sqrt{2}}  \\
0 & \frac{1}{\sqrt{2}} & 0 \\
\end{pmatrix},~
S_y = 
\begin{pmatrix}
0 & -\frac{i}{\sqrt{2}} & 0 \\
\frac{i}{\sqrt{2}} & 0 & -\frac{i}{\sqrt{2}}  \\
0 & \frac{i}{\sqrt{2}} & 0 \\
\end{pmatrix},~
S_z = 
\begin{pmatrix}
1 & 0 & 0 \\
0 & 0 & 0 \\
0 & 0 & -1\\
\end{pmatrix}.
\end{aligned}}
\label{eq:eq12}
\end{equation}
It is straightforward to obtain $\text{Tr} (\rho S_x) = \text{Tr} (\rho S_z) = 0$, and $\text{Tr} (\rho S_y)  = -2\sqrt{2} \text{Im}(b_1)$. 

The SDM or the spin-parity of $R$, can be determined through an angular analysis of a decay such as $R\to D_1 D_2$.
However, if the decay process also conserves parity, the imaginary part of $\rho$ cannot be measured by the angular analysis, while the real part of it can be measured~\cite{Chung:1974fq}. Consequently, odd-rank polarizations $d_L$ are not measurable, whereas even-rank polarizations are unaffected. Taking the hyperon decay $\Lambda\to p\pi^-$ as an example, its differential decay width is proportional to $1+\alpha \boldsymbol{\mathcal{P}} \cdot \boldsymbol{n}$ where $\alpha$ is the decay constant related to the weak interaction and $\boldsymbol{n}$ is the unit vector coinciding with one of the two decay products in the $\Lambda$ rest frame.
Under parity reversion, $\boldsymbol{\mathcal{P}}$ remains unchanged because it is an axial vector, whereas $\boldsymbol{n}$ changes sign. Therefore, a nonzero value of $\alpha$ signals parity violation. In strong and electromagnetic decays where $\alpha=0$, the spin-polarization vector $\boldsymbol{\mathcal{P}}$ is not measurable. In fact, as demonstrated in Eq.~(\ref{eq:eq6}), all $d_L$'s with odd $L$'s are not measurable if the decay is parity conserving.

Since $\text{Im}(\rho)$ is not measurable for parity-conserving decays of $R$, the SDM in Eq.~(\ref{eq:eq2}) can be further simplified. An SDM parametrization for spin-2 using angular parameters is given below~\cite{DiSalvo:2020run}.

\begin{equation}
{\scriptsize
\begin{aligned}
\begin{pmatrix}
\frac{1}{2}\cos^2\alpha_1 & \frac{1}{4}\sin2\alpha_1\cos\alpha_2\cos\beta_1 & \frac{\sqrt{2}}{4}\sin2\alpha_1\sin\alpha_2\cos\beta_3 & \frac{1}{4}\sin2\alpha_1\cos\alpha_2\cos\beta_5 & \frac{1}{2}\cos^2\alpha_1\cos\beta_6 \\
\frac{1}{4}\sin2\alpha_1\cos\alpha_2\cos\beta_1 & \frac{1}{2}\sin^2\alpha_1\cos^2\alpha_2 & \frac{\sqrt{2}}{4}\sin^2\alpha_1\sin2\alpha_2\cos\beta_2 & \frac{1}{2}\sin^2\alpha_1\cos^2\alpha_2\cos\beta_4 & -\frac{1}{4}\sin2\alpha_1\cos\alpha_2\cos\beta_5 \\
\frac{\sqrt{2}}{4}\sin2\alpha_1\sin\alpha_2\cos\beta_3 & \frac{\sqrt{2}}{4}\sin^2\alpha_1\sin2\alpha_2\cos\beta_2 & \sin^2\alpha_1\sin^2\alpha_2 & -\frac{\sqrt{2}}{4}\sin^2\alpha_1\sin2\alpha_2\cos\beta_2 & \frac{\sqrt{2}}{4}\sin2\alpha_1\sin\alpha_2\cos\beta_3 \\
\frac{1}{4}\sin2\alpha_1\cos\alpha_2\cos\beta_5 & \frac{1}{2}\sin^2\alpha_1\cos^2\alpha_2\cos\beta_4 & -\frac{\sqrt{2}}{4}\sin^2\alpha_1\sin2\alpha_2\cos\beta_2 & \frac{1}{2}\sin^2\alpha_1\cos^2\alpha_2 & -\frac{1}{4}\sin2\alpha_1\cos\alpha_2\cos\beta_1 \\
\frac{1}{2}\cos^2\alpha_1\cos\beta_6 & -\frac{1}{4}\sin2\alpha_1\cos\alpha_2\cos\beta_5 & \frac{\sqrt{2}}{4}\sin2\alpha_1\sin\alpha_2\cos\beta_3 & -\frac{1}{4}\sin2\alpha_1\cos\alpha_2\cos\beta_1 & \frac{1}{2}\cos^2\alpha_1 \\
\end{pmatrix}
\end{aligned}}
\label{eq:eq13}
\end{equation}
where $0\leq\alpha_{1,2}\leq\frac{\pi}{2}$ and $0\leq\beta_{1,2,3,4,5,6}\leq\pi$ are 8 independent angular parameters. The SDM of this form respects all the constraints on an SDM, including the positive semidefiniteness $|\rho_{ij}|^2\leq\rho_{ii}\rho_{jj}$, which is ensured by the $\cos\beta_i$ factors in the off-diagonal elements. It is worthwhile to mention that although $\text{Im}(\rho)$ is not experimentally accessible in parity-conserving decays, setting it to zero in the fit reduces the number of free parameters and can facilitate the experimental determination of the SDM when the available statistics are limited.
Likewise, the SDM of a spin-1 particle can be expressed as \begin{equation}
{\small
\begin{pmatrix}
\frac{1}{2}\cos^2\alpha & \frac{\sqrt{2}}{4}\sin2\alpha\cos\beta_1 & \frac{1}{2}\cos^2\alpha\cos\beta_2 \\
\frac{\sqrt{2}}{4}\sin2\alpha\cos\beta_1 & \sin^2\alpha & -\frac{\sqrt{2}}{4}\sin2\alpha\cos\beta_1 \\
\frac{1}{2}\cos^2\alpha\cos\beta_2 & -\frac{\sqrt{2}}{4}\sin2\alpha\cos\beta_1 & \frac{1}{2}\cos^2\alpha \\
\end{pmatrix},}
\label{eq:eq14}
\end{equation}
where $0\leq\alpha\leq\frac{\pi}{2}$ and $0\leq\beta_{1,2}\leq\pi$ are 3 independent angular parameters.


In the helicity formalism, if $R$ has a spin $J$ and a third component $M$ along the quantization axis, and  decays into two particles, the decay amplitude can be expressed as~\cite{Chung:1971ri}

\begin{equation}
\begin{aligned}
\mathcal{M}_{\lambda\nu}^J(\theta,\phi;M) & = \left< \theta\phi\lambda\nu|JM\lambda\nu\right> \left< JM\lambda\nu |\mathcal{M}| JM \right> \\
& \propto F_{\lambda\nu}^J D_{M\delta}^{J*}(\phi,\theta,0),\\
\end{aligned}
\label{eq:eq21}
\end{equation}
where $\theta$ and $\phi$ are the polar and azimuthal angles of one decay product in the rest frame of $R$, $\lambda$ and $\nu$  are the helicities of $D_1$ and $D_2$ in the final state, respectively, $\delta=\lambda-\nu$, $D^J_{M\delta}$ is the Wigner D-function, and $F_{\lambda\nu}^J = \left< JM\lambda\nu |\mathcal{M}| JM \right>$. If parity is conserved, the matrix elements $F_{\lambda\nu}^J$ will have the relation

\begin{equation}
F_{\lambda\nu}^J = \eta_R \eta_1 \eta_2 (-1)^{J - s_1 - s_2} F_{-\lambda, -\nu}^J,
\label{eq:eq22}
\end{equation}
where $\eta_R$, $\eta_1$, $\eta_2$ are the intrinsic parities of $R$, $D_1$ and $D_2$; $s_1$, $s_2$ are the total spins of $D_1$ and $D_2$, respectively.
In the following, we present several examples of angular analyses involving SDMs, assuming parity conservation throughout. 
These examples can be used either to determine the SDM of $R$ when its quantum numbers of $J^P$ are known, or to determine $J^P$ when the SDM is known.

\section{Decays of $2^+ \to 0^- 0^-$}

This is the simplest case for a $J=2$ particle, and $\eta_R$ must be $+1$ for the matrix element to be non-zero. In fact, since there is only one matrix element $F^2_{00}$, it factors out of the differential decay width and can be absorbed into the overall normalization.
Together with the SDM, one has

\begin{equation}
\begin{aligned}
\frac{d^2\Gamma}{d\cos\theta d\phi} & \propto \sum_{mm^\prime} \rho_{mm^\prime} D_{m0}^{2*}(\phi,\theta,0) D_{m^\prime0}^{2}(\phi,\theta,0) \\
& \propto \sum_{mm^\prime} \rho_{mm^\prime} e^{i(m-m^\prime)\phi} d_{m0}^{2}(\theta) d_{m^\prime0}^{2}(\theta)\\
& \propto \sum_{mm^\prime} \rho_{mm^\prime} \cos\left[(m-m^\prime)\phi\right] d_{m0}^{2}(\theta) d_{m^\prime0}^{2}(\theta) \\
\end{aligned}
\label{eq:eq23}
\end{equation}
where $d^J_{mm^\prime}$ is the small Wigner $d$-function.
In the last line, we have used the fact that $\rho$ is real and symmetric, so that the imaginary terms cancel in the summation.

\section{Decays of $2^+ \to 1^- 0^-$}

Because parity is conserved, one should have $\eta_R = \eta_1 \eta_2 (-1)^L$ where $L$ is the orbital angular momentum quantum number of the final $D_1 D_2$ system. This means that $L$ should be even. The total spin is $S=1$, and the angular-momentum coupling condition $|L-S|\leq J\leq L+S$ must also be satisfied. All these considerations leave only the choice of $L=2$. The matrix element $F_{\lambda\nu}^J$ can be decomposed into different partial-wave amplitudes~\cite{Chung:1971ri}

\begin{equation}
F_{\lambda\nu}^J = \sum_{LS} \sqrt{\frac{2L+1}{2J+1}} \left(L0S\delta|J\delta\right) \left(s_1\lambda s_2 -\nu|S\delta\right) G^J_{LS}, 
\label{eq:eq24}
\end{equation}
where $G^J_{LS} = \left< JM LS |\mathcal{M}| JM \right>$. With $\lambda=\pm 1,0$ and $\nu=0$, the three matrix elements can be expressed as

\begin{equation}
F^2_{10} = - F^2_{-10} = -\frac{1}{\sqrt{2}} G^2_{21},~~F^2_{00} = 0.
\label{eq:eq25}
\end{equation}
Therefore, the differential decay width is

\begin{equation}
\begin{aligned}
\frac{d^2\Gamma}{d\cos\theta d\phi} & \propto \sum_{mm^\prime \lambda} \rho_{mm^\prime} \left|F_{\lambda 0}^2\right|^2 D_{m\lambda}^{2*}(\phi,\theta,0) D_{m^\prime \lambda}^{2}(\phi,\theta,0) \\
& \propto \sum_{mm^\prime \lambda} \rho_{mm^\prime} \left|F_{\lambda 0}^2\right|^2  e^{i(m-m^\prime)\phi} d_{m\lambda}^{2}(\theta) d_{m^\prime \lambda}^{2}(\theta)\\
& \propto \sum_{mm^\prime} \rho_{mm^\prime} \cos\left[(m-m^\prime)\phi\right] \cdot \left[ d_{m1}^{2}(\theta) d_{m^\prime 1}^{2}(\theta) + d_{m -1}^{2}(\theta) d_{m^\prime -1}^{2}(\theta)\right]
\end{aligned} 
\label{eq:eq26}
\end{equation}
Notice that the sum over the helicity $\lambda$ of $D_1$ is performed after squaring the amplitude, since this helicity is assumed to be not measured.

\section{Decays of $2^+ \to 1^- 1^-$}

This is a more complicated case because partial waves with $L=0,2,4$ are all possible. Here we consider the simpler case in which the two daughter particles are identical.
Then, in addition to Eq.~(\ref{eq:eq22}), we have the constraint

\begin{equation}
F_{\lambda\nu}^J = (-1)^J F_{\nu\lambda}^J.
\label{eq:eq27}
\end{equation}
In the case of identical particles, the final state should also be symmetrized. Namely,

\begin{equation}
\mathcal{M}_{\lambda\nu}^J(\theta,\phi;M)_s = \tensor[_s]{ \left< \theta\phi\lambda\nu|JM\lambda\nu\right> }{_s} \tensor[_s]{\left< JM\lambda\nu |\mathcal{M}| JM \right> }{}
\label{eq:eq28}
\end{equation}
where the subscript $s$ means symmetrized states.  Explicitly,

\begin{equation}
\begin{aligned}
\ket{JM\lambda\nu}_s & = c_{\lambda\nu}\left[\ket{JM\lambda\nu} + (-1)^J \ket{JM\nu\lambda} \right], \\
\ket{\theta\phi\lambda\nu}_s & = c_{\lambda\nu}\left(\ket{\theta\phi\lambda\nu} + \ket{\pi-\theta,\phi+\pi,\nu\lambda} \right),
\end{aligned} 
\label{eq:eq29}
\end{equation}
where $c_{\lambda\nu} = 1/2$ ($1/\sqrt{2}$) for $\lambda=\nu$ ($\lambda\neq\nu$). It is interesting to note that Eq.~(\ref{eq:eq29}) also implies the Landau--Yang theorem: a $J=1$ particle cannot decay into two photons for which $\lambda=\nu=\pm 1$ must hold~\cite{Landau:1948kw,Yang:1950rg}. 
An expansion of the symmetrized two-particle helicity state, analogous to the usual case, can be written as:

\begin{equation}
\ket{\theta\phi\lambda\nu}_s  = \sum_{JM}\sqrt{\frac{2J+1}{4\pi}} D^J_{M\delta}(\phi,\theta,0) \ket{JM\lambda\nu}_s.
\label{eq:eq30}
\end{equation}
Therefore, the decay amplitude for $J=2$ is

\begin{equation}
\mathcal{M}_{\lambda\nu}^2(\theta,\phi;M)_s \propto c_{\lambda\nu} F_{\lambda\nu}^2  D_{M\delta}^{2*}(\phi,\theta,0)
\label{eq:eq31}
\end{equation}
The differential decay width is then

\begin{equation}
{\small
\begin{aligned}
\frac{d^2\Gamma}{d\cos\theta d\phi} \propto & \sum_{\stackanchor{$\scriptstyle mm^\prime \lambda\nu$}{$\scriptstyle (\lambda\geq\nu)$} } \rho_{mm^\prime} c_{\lambda \nu}^2 \left|F_{\lambda \nu}^2\right|^2 D_{m\delta}^{2*}(\phi,\theta,0) D_{m^\prime \delta}^{2}(\phi,\theta,0) \\
\propto & \sum_{mm^\prime} \rho_{mm^\prime} \cos\left[(m-m^\prime)\phi\right] \cdot \left[ \left( \frac{1}{2} \left|F_{11}^2\right|^2 + \frac{1}{4} \left|F_{00}^2\right|^2 \right) d_{m0}^{2}(\theta) d_{m^\prime 0}^{2}(\theta) \right. \\
& \left. + \left|F_{10}^2\right|^2  d_{m1}^{2}(\theta) d_{m^\prime 1}^{2}(\theta) + \frac{1}{2} \left|F_{1-1}^2\right|^2  d_{m2}^{2}(\theta) d_{m^\prime 2}^{2}(\theta) \right], \\
\end{aligned}}
\label{eq:eq32}
\end{equation}
where $F_{11}^2$, $F_{00}^2$, $F_{10}^2$ and $F_{1-1}^2$ are four independent parameters. 
The remaining matrix elements are determined by these four and therefore do not appear independently. 
To distinguish $F_{11}^2$ from $F_{00}^2$, the correlation between the $D_1$ and $D_2$ decays should also be taken into account in the angular analysis. However, measuring both $\rho$ and $F_{\lambda\nu}^2$ by fitting to experimental data is difficult. 
In practice, one typically needs to impose theoretically motivated assumptions on one set of parameters in order to extract the other.

\section{Decays of $1^+ \to 1^- 0^-$}

It is easy to see that $L=0,2$ in this case.  The matrix element $F_{\lambda\nu}^J$ can be decomposed as 

\begin{equation}
\begin{aligned}
 F^1_{10}  = F^1_{-10} = \frac{1}{\sqrt{3}} G^1_{01} + \frac{1}{\sqrt{6}} G^1_{21}, \\
 F^1_{00}  = \frac{1}{\sqrt{3}} G^1_{01} - \sqrt{\frac{2}{3}} G^1_{21}. \\
\end{aligned}
\label{eq:eq33}
\end{equation}
Two limiting scenarios can be considered based on theoretical input: S-wave dominance and D-wave dominance.
If the decay is S-wave dominated, then one has $|F^1_{10}|^2 = |F^1_{-10}|^2 = |F^1_{00}|^2$. The differential decay width reads

\begin{equation}
{\small
\begin{aligned}
\frac{d^2\Gamma}{d\cos\theta d\phi} & \propto \sum_{mm^\prime \lambda} \rho_{mm^\prime} \left|F_{\lambda 0}^1\right|^2 D_{m\lambda}^{1*}(\phi,\theta,0) D_{m^\prime \lambda}^{1}(\phi,\theta,0) \\
& \propto \sum_{mm^\prime} \rho_{mm^\prime} \cos\left[(m-m^\prime)\phi\right] \cdot \left[ d_{m1}^{1}(\theta) d_{m^\prime 1}^{1}(\theta) + d_{m0}^{1}(\theta) d_{m^\prime 0}^{1}(\theta) + d_{m -1}^{1}(\theta) d_{m^\prime -1}^{1}(\theta)\right].
\end{aligned}}
\label{eq:eq34}
\end{equation}
In the case of a D-wave-dominated process, one has $|F^1_{10}|^2 = |F^1_{-10}|^2 = \frac{|F^1_{00}|^2}{4}$, and the corresponding differential decay width reads

\begin{equation}
\small{
\frac{d^2\Gamma}{d\cos\theta d\phi} \propto \sum_{mm^\prime} \rho_{mm^\prime} \cos\left[(m-m^\prime)\phi\right] \cdot \left[ d_{m1}^{1}(\theta) d_{m^\prime 1}^{1}(\theta) + 4d_{m0}^{1}(\theta) d_{m^\prime 0}^{1}(\theta) + d_{m -1}^{1}(\theta) d_{m^\prime -1}^{1}(\theta)\right]. }
\label{eq:eq35}
\end{equation}

In summary, the SDM contains complete information about a particle’s spin state, and may provide important information about its production mechanism.
It can be measured through angular analyses of its decay products.
The number of independent parameters in an SDM can be significantly reduced when the production process conserves parity, as is typically the case for the strong production in hadron-hadron collisions and the electromagnetic production electron-positron collisions. It will greatly facilitate the measurement of the SDM elements by fitting to experimental data. However, if the particle decay also conserves parity, the imaginary part of SDM cannot be measured through angular analysis. In fact, the imaginary (real) part of SDM is related to odd-rank (even-rank) polarizations. Specific parameterizations for spin-1 and spin-2 SDMs are provided in the helicity frame, which satisfy all the physical constraints expected of an SDM. Examples of differential decay widths for several spin-parity hypotheses with parity conservation are also provided in conjunction with the SDM.
Both the SDM parameters and the independent decay matrix elements enter the angular distributions. One can extract the values of one set by assuming fixed values of the other and fitting to some experimental data. When the final state particles have low spins, there are fewer independent matrix elements, and the SDM may be determined in a less model-dependent way. Combining multiple decay modes or sequential decays may improve the prospects for simultaneously determining both the SDM and the decay matrix elements.

\printbibliography

\end{document}